\UseRawInputEncoding
\documentclass[showpacs,twocolumn,superscriptaddress,nofootinbib]{revtex4-1} 

\usepackage{hyperref}
\hypersetup{colorlinks=true,linkcolor=blue,citecolor=cyan}
\usepackage{graphicx}
\usepackage{xcolor, soul}
\sethlcolor{green}
\usepackage{dcolumn}
\usepackage{bm}
\usepackage{color}
\usepackage{enumitem}
\usepackage{mathrsfs}
\usepackage{amsmath}
\usepackage{amssymb}
\usepackage{cleveref}
\usepackage{orcidlink}

\crefname{equation}{Eqn.}{Eqns.}
\crefname{figure}{Fig.}{Figs.}
\crefname{section}{Sec.}{Sec.}
\crefname{table}{Table}{Tables}
\usepackage{physics}
\usepackage{multirow}

\setstcolor{red}

\begin{document}

\title{{Probing the Schwarzschild black hole immersed in a dark matter halo through astrophysical tests 
}}

\author{Tursunali Xamidov}
\email{xamidovtursunali@gmail.com}
\affiliation{Institute of Fundamental and Applied Research, National Research University TIIAME, Kori Niyoziy 39, Tashkent 100000, Uzbekistan} 
\affiliation{Tashkent State Technical University, 100095 Tashkent, Uzbekistan}

\author{Sanjar Shaymatov}
\email{sanjar@astrin.uz}
\affiliation{Institute of Fundamental and Applied Research, National Research University TIIAME, Kori Niyoziy 39, Tashkent 100000, Uzbekistan}
\affiliation{Institute for Theoretical Physics \& Cosmology, Zhejiang University of Technology, Hangzhou 310023, China}
\affiliation{University of Tashkent for Applied Sciences, Str. Gavhar 1, Tashkent 100149, Uzbekistan}
\affiliation{Western Caspian University, Baku AZ1001, Azerbaijan}

\author{Qiang Wu}
\email{wuq@zjut.edu.cn}
\affiliation{Institute for Theoretical Physics \& Cosmology, Zhejiang University of Technology, Hangzhou 310023, China}
\affiliation{United Center for Gravitational Wave Physics (UCGWP), Zhejiang University of Technology, Hangzhou 310023, China}

\author{Tao Zhu}
\email{zhut05@zjut.edu.cn}
\affiliation{Institute for Theoretical Physics \& Cosmology, Zhejiang University of Technology, Hangzhou 310023, China}
\affiliation{United Center for Gravitational Wave Physics (UCGWP), Zhejiang University of Technology, Hangzhou 310023, China}

\date{\today}
\begin{abstract}

We investigate a recently derived Schwarzschild-like black hole immersed in a Dehnen-type $(\alpha,\beta,\gamma)=(1,4,5/2)$ dark matter (DM) halo. We obtain constraints on the two model parameters, i.e., the halo core radius $r_s$ and the DM density parameter $\rho_s$ in both the weak and the strong field regimes.  
In the weak field, we model test particle geodesics and match the predicted perihelion shift to Mercury (Solar System) and the orbit of the S2 star data. We obtain upper limits on $r_s$ and $\rho_s$ and highlight that the DM halo effects become observable only around supermassive BHs.  
In the strong field, we analyse twin high frequency quasiperiodic oscillations (QPOs) from four microquasars (e.g., GRO~J1655-40, GRS~1915+105, XTE~J1859+226, and XTE~J1550-564). Because QPO frequencies depend only on the local spacetime curvature, they can serve as a probe of halo-induced deviations from general relativity. Our MCMC analysis produces posterior distributions for model parameters, revealing close agreement between the theoretical QPO frequencies and the observations for GRS 1915+105 and GRO J1655-40. The same analysis also yielded best-fit values and upper bounds for each parameter. 
Our combined geodesic and QPO analysis demonstrates that timelike orbits and epicyclic oscillations can act as sensitive probes of DM halos around BHs, offering a pathway to distinguish Dehnen-type profiles from alternative DM distributions in future analysis and observations. 

\end{abstract}

\maketitle

\section{Introduction}
\label{introduction}

Black hole (BH) astrophysics has entered a precision era. Horizon-scale images of M87$^{\star}$ and Sgr A$^{\star}$ captured by the Event Horizon Telescope \cite{Akiyama19L1,Akiyama19L6}, together with the steadily growing set of gravitational wave (GW) detections~\cite{Abbott16a,Abbott16b}, now allow the strong-gravity regime to be studied with direct data rather than purely through theory.  In theoretical models, black holes are often treated as isolated objects, yet in reality they continuously interact with the surrounding baryonic medium. Astronomical observations further reveal an additional, non-baryonic component - dark matter - that accounts for roughly 27 \% of the Universe's total mass-energy content, and thus also plays a crucial role in cosmic structure and evolution~\cite{Aghanim2018PC}. Because DM neither emits, absorbs, nor scatters electromagnetic radiation, it remains invisible and can be traced only through the gravity it exerts. Its fingerprints first emerged in the galactic rotation curves, where stars orbit far faster than luminous matter alone can support \cite{Rubin70ApJ}. Similar mass discrepancies appear in the gravitational lensing maps of galaxy clusters, whose deflection angles exceed general-relativistic predictions based on visible mass \cite{Karamazov21ApJ,Qi23}.

Although plenty of evidence supports the existence of dark matter (DM), its non-gravitational properties remain unknown. Various theoretical models have been proposed: some treat DM as weakly interacting massive particles (WIMPs) coupling via gravity and the weak nuclear force, while others invoke alternative candidates such as axions or sterile neutrinos \cite{Boehm04NPB,Bertone05PhR,Feng09JCAP,Schumann19}. Because the influence of DM is very weak, we can detect it only in very massive systems. Therefore, studying DM in strong gravitational fields is extremely important. In most models of a black hole-dark matter system, the DM is treated as a halo surrounding the black hole \cite{GondoloPRL99, BERTONEMPL05}. One widely used example is the Dehnen type halo model. Introduced by Walter Dehnen in 1993 to represent stellar bulges and DM halos, this model is defined by a three-parameter family of spherically symmetric density profiles, specified by the parameters $(\alpha, \beta, \gamma)$~\cite{Dehnen93}. Later on, it was developed in Refs.~\cite{Matos05,Xu18JCAP}. Recently, within the Dehnen-type DM halo model, new Schwarzschild-like BH solutions embedded in a DM halo were derived for Dehnen-type density profiles (e.g., see \cite{Gohain24DM,ShaymatovDM,Uktamov25DM}). Within the density profile of the Dehnen-type DM halo, recent investigations explore the impact of the DM halo on various astrophysical processes, including quasinormal modes, the BH shadow, and gravitational waveforms associated with periodic orbits \cite{Al-Badawi25CPC,Al-Badawi25CTP_DM,Alloqulov-Xamidov25,Arpan2025PhRvD}.

High-frequency quasiperiodic oscillations (HF QPOs) in the X-ray flux  coming from binary systems (e.g., BH and neutron star) provide a sensitive tool for probing the strong gravity region close to compact objects. QPOs appear as narrow peaks in the X-ray power-density spectrum. They originate in the inner accretion disks of compact objects such as black holes and neutron stars, and their frequencies depend only the curvature of the spacetime. Therefore, HF QPO measurements have been widely used to test alternative gravity models by analyzing X-ray data from accretion disks \cite{Abramowicz13,Bambi12a,Bambi16b,Tripathi19,Xamidov25PDU,Orlando2022ApJ}. 

However, magnetic fields can also influence the motion of slightly charged matter orbiting a magnetized black hole, thereby affecting the QPOs formed in the accretion disk \cite{Stuchlik2020Univ}.  Moreover, there are works that systematically study the influence of a wide variety of rotating black hole spacetimes, including those modified by dark matter halos and alternative gravity scenarios, on the properties of HF QPOs \cite{Shahzadi2024CQG}.

HF QPOs commonly appear as a pair of peaks whose upper $\nu_U$ and lower $\nu_L$ frequencies maintain an approximate $3:2$ ratio. Such twin peak HF QPOs have been observed in multiple Galactic microquasars (e.g., GRO J1655-40, GRS 1915+105, XTE J1859+226, and XTE J1550-564) \cite{Kluzniak01,Torok05A&A,Remillard06ApJ}.
Many models interpret these QPO peaks as signatures of relativistic epicyclic motion or disk-oscillation resonances near the innermost stable circular orbit (ISCO) \cite{Stuchlik13A&A,Stella99-qpo,Rezzolla_qpo_03a}. Despite ongoing debate over the precise mechanism behind their generation \cite{Torok11A&A}, a range of theories have been proposed, including models that incorporate magnetic fields around black holes \cite{Tursunov20ApJ,Panis19,Shaymatov20egb,Shaymatov22c}. Recent studies continue to evaluate and refine these explanations, exploring various resonance conditions and alternative spacetime metrics (see, e.g., \cite{Germana18qpo,Tarnopolski:2021ula,Dokuchaev:2015ghx,Kolos15qpo,Aliev12qpo,Stuchlik07qpo,Titarchuk05qpo,Azreg-Ainou20qpo,Jusufi21qpo,Ghasemi-Nodehi20qpo,Rayimbaev22qpo,Shaymatov23ApJ,Shaymatov23qpo,Mustafa24PDU1,Liu:2023vfh,Guo:2025zca,Liu:2023ggz,Yan_2023}).

In this work, we investigate the solution of the BH-Dehnen-type DM halo and explore its impact in both weak and strong field regimes and derive the corresponding constraints on the parameters $r_s$ and $\rho_s$. First, we study the motion of particles based on the principles of general relativity and derive the weak-field limit by matching the model to observational data from Mercury and the orbit of the S2 star. To constrain the model parameters $r_s$ and $\rho_s$ in strong gravitational fields, we turn to quasiperiodic oscillations. Because QPO frequencies depend only on the space-time curvature around black holes, they provide a powerful probe of general relativity and of possible dark-matter halo effects in the strong-field regime. Accordingly, we use QPO data from the microquasars GRO J1655-40, GRS 1915+105, XTE J1859+226, and XTE J1550-564 to set strong field limits on the parameters $r_s$ and $\rho_s$.

The paper is structured as follows. In Sec.~\ref{Sec:metric}, we introduce the spacetime metric describing a Schwarzschild black hole embedded in a dark matter halo and analyze particle dynamics. We also define the weak-field parameter space for the DM parameters $r_s$ and $\rho_s$ by examining the perihelion precession of Mercury and the orbit of the S2 star. In Sec.~\ref{Section:EpyFreq}, we explore the dynamics of epicyclic motion and derive general expressions for the upper and lower frequencies, while assessing how variations in the DM halo parameters influence these frequencies. In Sec.~\ref{Sec:MCMC}, we employ a Markov Chain Monte Carlo (MCMC) method to place strong-field constraints on $r_s$ and $\rho_s$ using observational data from four QPO sources. The paper concludes with a summary in Sec.~\ref{Sec:conclusion}. We use geometric units $G=c=1$ throughout this work. 

\section{The spacetime metric and the dynamics of motion}\label{Sec:metric}

The Dehnen dark matter halo density profile arises as a special case of the double power-law profile for $(\alpha, \beta, \gamma) = (1,4,\gamma)$ \cite{MoHoujunBook2010}:
\begin{equation}
    \rho = \rho_{s} \left( \frac{r}{r_{s}} \right)^{-\gamma}
    \left[ \left( \frac{r}{r_{s}} \right)^{\alpha} + 1 \right]^{\frac{\gamma - \beta}{\alpha}} ,
\end{equation}
where $\rho_{s}$ and $r_{s}$ denote the characteristic halo density and scale radius, respectively.  
The parameter $\gamma$ takes values in the interval $[0,3]$.  
The corresponding mass profile is obtained from
\begin{equation}
    M_{D} = 4\pi \int_{0}^{r} \rho(r')\, r'^{2} dr'\, .
\end{equation}
In this work, we focus on a black hole solution with a DM halo within a Dehnen type density profile $(\alpha,\beta,\gamma) = (1,4,5/2)$ DM halo.
The spacetime metric of a Schwarzschild black hole embedded in the Dehnen type DM halo (1,4,5/2) can be written as (see Ref.~\cite{ShaymatovDM} for details):
\begin{equation} \label{spacetime}
ds^2 = -f(r) dt^2 + f(r)^{-1} dr^2 + r^2 ( d\theta^2 + \sin^2 \theta d\phi^2)\, ,
\end{equation}
where 
\begin{equation}
    f(r)=1-\frac{2M}{r}-32\pi \rho_s r^3_s\sqrt{\frac{r+r_s}{r^2_s r}} \, .
\end{equation}

Suppose that a neutral particle is moving around a Schwarzschild black hole embedded in a Dehnen-type dark matter halo. Then the Hamiltonian describing the massive particle’s motion can be written as follows \cite{Misner73}:
\begin{equation}\label{hamiltonian}
H=\frac{1}{2}g^{\mu\nu}p_\mu p_\nu = -\frac{m^2}{2}\ ,
\end{equation}
where $p^\mu$ denotes the four‐momentum of the neutral particle. The indices $\mu$ and $\nu$ run over $(t,\,r,\,\theta,\,\phi)$ in the spherical coordinates. One can write the four-momentum as $p^\mu \;=\; m\,u^\mu$, where $m$ is the particle’s rest mass and $u^\mu \;=\; dx^\mu/d\tau$ is its four-velocity, with $\tau$ denoting proper time. Since the metric in equation \eqref{spacetime} does not depend explicitly on $t$ or $\phi$, one immediately obtains the following conserved quantities:
\begin{equation} \label{conservation-E}
    \frac{p_t}{m}=g_{tt} u^t=g_{tt}\frac{dt}{d\tau}=-\mathcal{E}\ ,
\end{equation}
\begin{equation} \label{conservation-L}
\frac{p_\phi}{m}=g_{\phi\phi} u^\phi=g_{\phi\phi}\frac{d\phi}{d\tau}=\mathcal{L}\ ,
\end{equation}
where $\mathcal{E}$ and $\mathcal{L}$ are the specific energy and the specific angular momentum of the particle. 

Using Eqs.~(\ref{hamiltonian}-\ref{conservation-L}), we can write the following equation:
\begin{align} \label{ex-hamiltonian}
   \frac{1}{2}\left(g^{rr}p_r^2+ g^{\theta\theta}p_\theta^2\right) = -m^2\,H_{pot}  \, ,
\end{align}
where
\begin{equation} \label{Hpot}
    H_{pot} = \frac{1}{2} \left( g^{tt} \mathcal{E}^2 + g^{\phi\phi} \mathcal{L}^2 + 1 \right)\, .
\end{equation}
Together with the relation $p_\mu = g_{\mu\nu} p^\nu$, we rewrite Eq.~\eqref{ex-hamiltonian} as
\begin{equation} \label{eqofmotion}
    g_{rr}\dot{r}^2+ g_{\theta\theta}\dot{\theta}^2 = -2H_{pot} \ ,
\end{equation}
where the over-dot denotes differentiation with respect to the proper time $\tau$.

Assuming the particle moves in the equatorial plane ($\theta = \pi/2 = \text{const}$), so that $\dot\theta = 0$, the second term in Eq.~\eqref{eqofmotion} vanishes and the equation reduces to
\begin{equation}  \label{equatorial-motion}
    g_{rr}\,\dot r^2 = -2H_{pot}\,.
\end{equation}

Combining Eq.~\eqref{conservation-L} with Eq.~\eqref{equatorial-motion} gives the trajectory equation
\begin{eqnarray} \label{equatorial-motion2}
    \left(\frac{dr}{d\phi}\right)^2 = -\frac{2g_{\phi\phi}^2 H_{pot}}{g_{rr}\mathcal{L}^2} \, .
\end{eqnarray}
One can rewrite Eq.~\eqref{equatorial-motion2} in the following form using Eq.~\eqref{Hpot}:
\begin{eqnarray} \label{equatorial-motion3}
    \left(\frac{dr}{d\phi}\right)^2 = \frac{r^4}{\mathcal{L}^2}\left[\mathcal{E}^2 - f(r)\left(1+\frac{\mathcal{L}^2}{r^2}\right)\right] \, .
\end{eqnarray}

For a circular orbit with $r=\text{const}$ and $\dot r = 0$, lying in a plane through the origin that is inclined at some angle to the equatorial plane, Eq.~\eqref{eqofmotion} reduces to
\begin{equation} \label{latitudinal-motion}
g_{\theta\theta}\dot\theta^2 = -2H_{\text{pot}} \, .
\end{equation}
Using Eq.~\eqref{conservation-L} together with Eq.~\eqref{latitudinal-motion}, one obtains the trajectory equation
\begin{eqnarray} \label{equatorial-motion2}
\left(\frac{d\theta}{d\phi}\right)^2 = -\frac{2g_{\phi\phi}^2 H_{\text{pot}}}{g_{\theta\theta}\mathcal{L}^2} \, .
\end{eqnarray}

For a particle on a circular orbit of radius $R$ in the equatorial plane ($\theta = \pi/2$), the following conditions hold:
\begin{equation} \label{conditions}
\begin{aligned}
    H_{\text{pot}}(R, \frac{\pi}{2}) &= 0 \, , \\
    \left. \partial_r H_{\text{pot}}(r,\theta) \right|_{r = R} &= 0 \, , \\
    \left. \partial_\theta H_{\text{pot}}(r,\theta) \right|_{\theta = \frac{\pi}{2}} &= 0 \, .
\end{aligned}
\end{equation}

\begin{figure*}
    \centering
    \includegraphics[scale=0.5]{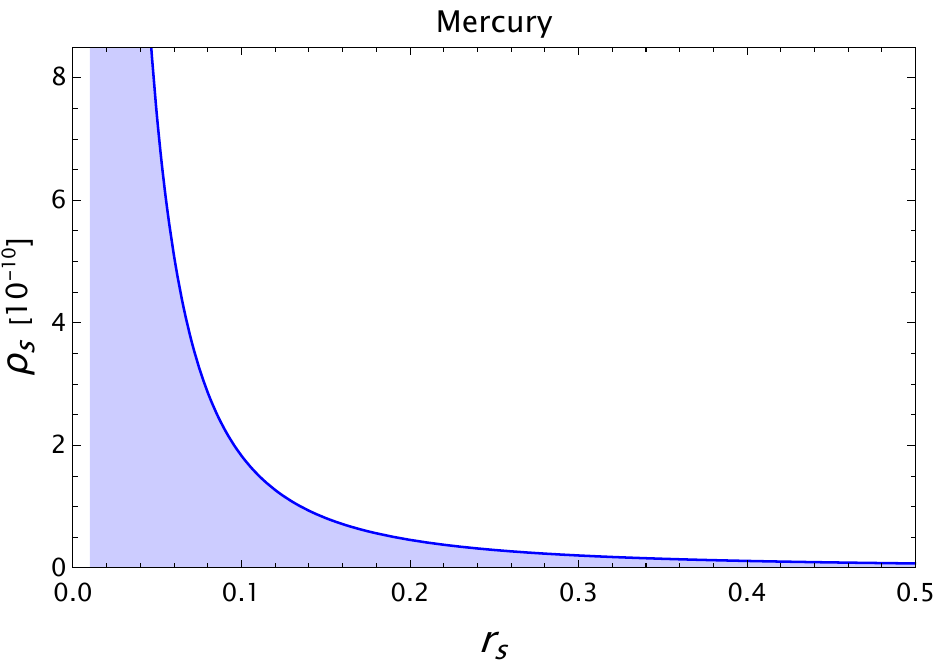}
    \includegraphics[scale=0.5]{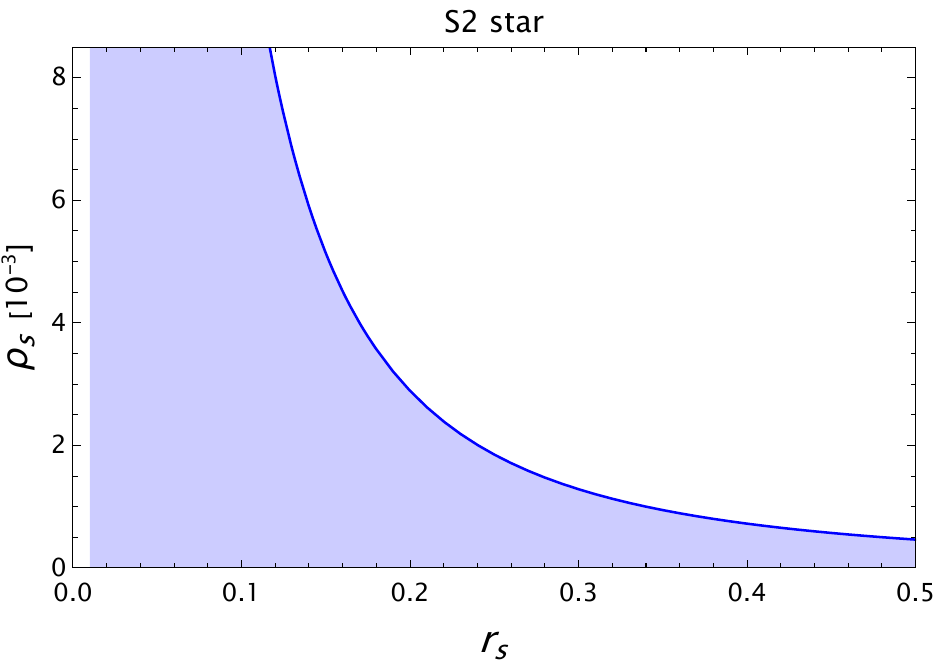}
    \caption{Parameter space between the dimensionless halo radius $r_{s}$ and the dimensionless dark‐matter density $\rho_{s}$. The shaded region indicates the observationally allowed values of $(r_{s},\,\rho_{s})$ from Mercury (left) and S2 star (right) measurements. 
 }
    \label{fig:parspace}
\end{figure*}

From Eqs.~\eqref{conservation-E} and~\eqref{conservation-L}, one can determine the following expression for the angular velocity of the particle as measured by a distant observer:
\begin{equation}\label{omega1}
    \frac{d\phi}{dt} = \Omega = -\frac{g_{tt}}{g_{\phi\phi}} \cdot \frac{\mathcal{L}}{\mathcal{E}} \, .
\end{equation}
On the other hand, as shown in Refs.~\cite{Shaymatov22a,Shaymatov23qpo}, the angular velocity can also be expressed as:
\begin{equation} \label{omega2}
    \Omega = \pm \sqrt{ -\frac{g_{tt,r}}{g_{\phi\phi,r}}} \, .
\end{equation}

$\Omega$ for equatorial ($\theta=\pi/2$) circular motion can be determined from the metric in Eq.~\eqref{spacetime} using the above relation, yielding
\begin{equation}\label{omega3}
\Omega = \pm \sqrt{\frac{M}{r^3}+\frac{8\pi\rho_s r_s^3}{r^3}\sqrt{\frac{r}{r+r_s}}} \ .
\end{equation}

By solving Eq.~\eqref{Hpot} under the condition given in Eq.~\eqref{conditions} and considering the equality in Eq.~\eqref{omega1}, we obtain the following expressions for the energy and angular momentum:
\begin{equation} \label{spec-energy}
\mathcal{E} = -\frac{g_{tt}}{\sqrt{-g_{tt} - g_{\phi\phi}\Omega^2}},
\end{equation}
\begin{equation}\label{spec-ang-momentum}
\mathcal{L} = \frac{g_{\phi\phi}\Omega}{\sqrt{-g_{tt} - g_{\phi\phi}\Omega^2}} \ .
\end{equation}
Using Eqs.\eqref{spec-energy} and \eqref{conservation-E}, we can write the following expression for $u^t$
\begin{equation}\label{four-vel}
u^t = \frac{1}{\sqrt{-g_{tt} - g_{\phi\phi}\Omega^2}} \ ,
\end{equation}
The expressions for $u^t$, $\mathcal{E}$, and $\mathcal{L}$ can be obtained using the metric in Eq.~\eqref{spacetime} and the angular velocity defined in Eq.~\eqref{omega3}
\begin{equation}\label{four-vel-exp}
u^t = \left(1-\frac{3 M}{r}-\frac{8 \pi\rho_s r_s^2 (4 r+5 r_s)}{\sqrt{r (r+r_s)}}\right)^{-\frac{1}{2}} \ ,
\end{equation}
\begin{equation} \label{spec-energy-exp}
\mathcal{E} =\left(1-\frac{2M}{r}-32\pi \rho_s r^3_s\sqrt{\frac{r+r_s}{r^2_s r}}\right)\times u^t,
\end{equation}

\begin{equation}\label{spec-ang-momentum-exp}
\mathcal{L} = \sqrt{ r\left(M +8 \pi  \rho_s r_s^3 \sqrt{\frac{r}{r+r_s}}\right)\sin ^2(\theta )} \times u^t\, .
\end{equation}
Let us perform the following transformation:
\begin{equation}
    \xi = \frac{1}{r} , \quad \frac{d\xi}{d\phi} = -\frac{1}{r^2}\frac{dr}{d\phi} \, .
\end{equation} 
As a result, Eq.~\eqref{equatorial-motion3} takes the form
\begin{equation} \label{eqGeodesic}
    \left(\frac{d\xi}{d\phi}\right)^2 = \frac{\mathcal{E}^2}{\mathcal{L}^2} - \frac{f(\xi)}{\mathcal{L}^2} \left(1 + \mathcal{L}^2 \xi^2\right) \, ,
\end{equation}
where the function $f(\xi)$ is given by
\begin{equation}
    f(\xi) = 1 - 2 M \xi - 32\pi \rho_s r_s^3 \sqrt{\frac{1 + r_s \xi}{r_s^2}} \, .
\end{equation}

Differentiating Eq.~\eqref{eqGeodesic} with respect to $\phi$ yields the geodesic equation for a massive test particle.
\begin{align} \label{orbital-equation2}
    &\frac{d^2 \xi}{d\phi^2} =  \frac{M}{\mathcal{L}^2}- \xi + \frac{g(\xi)}{\mathcal{L}^2} \, , 
\end{align}
where
\begin{equation}\label{gf}
    \frac{g(\xi)}{\mathcal{L}^2} = 3 M \xi^2 
    +   \frac{8\pi\rho_s r_s^2}{\sqrt{1+r_s \xi}}\left(\frac{r_s}{\mathcal{L}^2}+4\xi+5r_s \xi^2\right)\, .
\end{equation}
Following the approach outlined in Ref.~\cite{Adkins_2007}, the perihelion shift after a complete revolution is defined as:
\begin{equation}
    \Delta \phi = \frac{\pi}{\mathcal{L}^2} \left| \frac{d g(\xi)}{d\xi} \right|_{\xi = \frac{1}{b}}\, .
\end{equation}
Using Eq.~\eqref{gf}, we can determine the perihelion shift as follows
\begin{eqnarray}\label{pershift}
     \Delta \phi &=& \frac{6 \pi  M}{\mathcal{R}}+4 \pi ^2 \rho_s r_s^2 \sqrt{\frac{\mathcal{R}+r_s}{\mathcal{R}}}\cdot\frac{ 8 \mathcal{R}^2+24 \mathcal{R} r_s+15 r_s^2}{(\mathcal{R}+r_s)^2}\nonumber\\&-&\frac{4 \pi ^2 \rho_s r_s^4}{\mathcal{L}^2 }\left(\frac{\mathcal{R}}{\mathcal{R}+r_s}\right)^{3/2}\, ,
\end{eqnarray}
where we have defined $\mathcal{R} = a(1 - e^2)$ with $a$ and $e$ representing the semi-major axis and eccentricity of the orbit, respectively. 

To recover physical dimensions, the mass $M$, the
angular momentum $L^2$, the dark matter halo density $\rho_s$, and the halo radius $r_s$ are rescaled as
\begin{eqnarray}
M  &\Rightarrow& GM/c^2 \, ,\\
L^2 &\Rightarrow& GMa(1 - e^2)/c^2 \, ,\\
\rho_s &\Rightarrow& \frac{3c^4}{32\pi G^2M^2}\rho_s \, ,\\
r_s&\Rightarrow&\frac{G M}{c^2}r_s \, .
\end{eqnarray}

The dark matter halo density $\rho_{s}$ and the halo radius $r_{s}$ on the right-hand side are therefore
\textit{dimensionless}: they give the numerical values of the corresponding
physical quantities when expressed in the units
$3c^{4}/(32\pi G^{2}M^{2})$ and $GM/c^{2}$, respectively.

With these conventions, the perihelion shift after one full revolution,
Eq.~\eqref{pershift}, becomes
\begin{eqnarray} \label{eq:perishift}
    &&\Delta\phi = 6\pi\alpha +3 \pi  \rho_s r_s^2 \sqrt{1+\alpha r_s}\bigg(1+\frac{\alpha r_s}{1+\alpha r_s}-\nonumber\\ &-& \frac{\alpha r^2_s}{8(1+\alpha r_s)}-\frac{\alpha^2 r^2_s}{8(1+\alpha r_s)^2}+\frac{\alpha^2 r^3_s}{8(1+\alpha r_s)^2}\bigg)\, ,
\end{eqnarray}
where
\begin{equation}
    \alpha = \frac{GM}{a c^2 (1-e^2)}\, .
\end{equation}
Here, $\rho_s$ and $r_s$ denote the dimensionless dark matter halo density and the halo radius, respectively.

We use observational data from the perihelion shifts of Mercury and the S2 star to determine constraints on $r_s$ and $\rho_s$. Below, we present the observational data for the orbital parameters and the perihelion shift of Mercury and the S2 star \cite{Benczik02PRD,Iorio15IJMPD,Iorio2019ApJ,Shaymatov23ApJ,AbuterAmorim2020}. 

\vspace{1em}
\noindent\textbf{Mercury:}
\begin{eqnarray*}
    \frac{2 G M_\odot}{c^2} &=& 2.95325008 \times 10^3 \, [\text{m}] \, , \nonumber \\
    a &=&  5.7909175 \times 10^{10} \, [\text{m}] \, , \nonumber \\
    e &=&  0.20563069\, , \nonumber \\
    \Delta \phi_{\text{obs}} &=& 2\pi \times (7.98734 \pm 0.00037) \times 10^{-8}~\mathrm{rad/rev}.
\end{eqnarray*}
\vspace{1em}
\noindent\textbf{S2 star:}
\begin{eqnarray*}
    M_{\text{Sgr A}^*} &=& 4.260 \times 10^6 M_{\odot}\, , \\
    a_{\text{S2}} &=& 970 \, [\text{au}] \, , \\
    1 \ \text{au} &=& 1.495978707\times10^{11} \, [\text{m}]\, , \\
    e_{\text{S2}} &=& 0.884649 \, , \\
    T_{\text{S2}} &=& 16.052 \, [\text{years}] \, , \\
    \Delta\phi_{\text{obs}} &=& 48.298 \ f_{\text{SP}} \ \Big[\ ^{\prime\prime}/\text{year} \Big] \, ,\, f_{\text{SP}} = 1.10 \pm 0.19  \, .
\end{eqnarray*}

Based on these data, we use Eq.~\eqref{eq:perishift} to plot the parameter space of $r_s$ and $\rho_s$ for both Mercury and the S2 star (see Fig.~\ref{fig:parspace}). From Fig.~\ref{fig:parspace}, it is evident that the parameter space for the S2 star is larger than that for Mercury. This indicates that the influence of the DM halo becomes more evident near more massive objects. It is evident that the scale of $\rho_s$ is of the order of about $10^{-10}$ in the weak field and its ratio accordingly refers to the order of $10^{-7}$ between the solar system and the supermassive BH at the center of the $\text{Sgr A}^*$ galaxy, similar to the ratio $M_\odot/M_{\text{Sgr A}^*}$. This highlights that the effects of the DM halo become observable only around supermassive BHs.

\section{The dynamics of epicyclic motion}
\label{Section:EpyFreq}

\begin{figure*}
    \centering
    \includegraphics[scale=0.5]{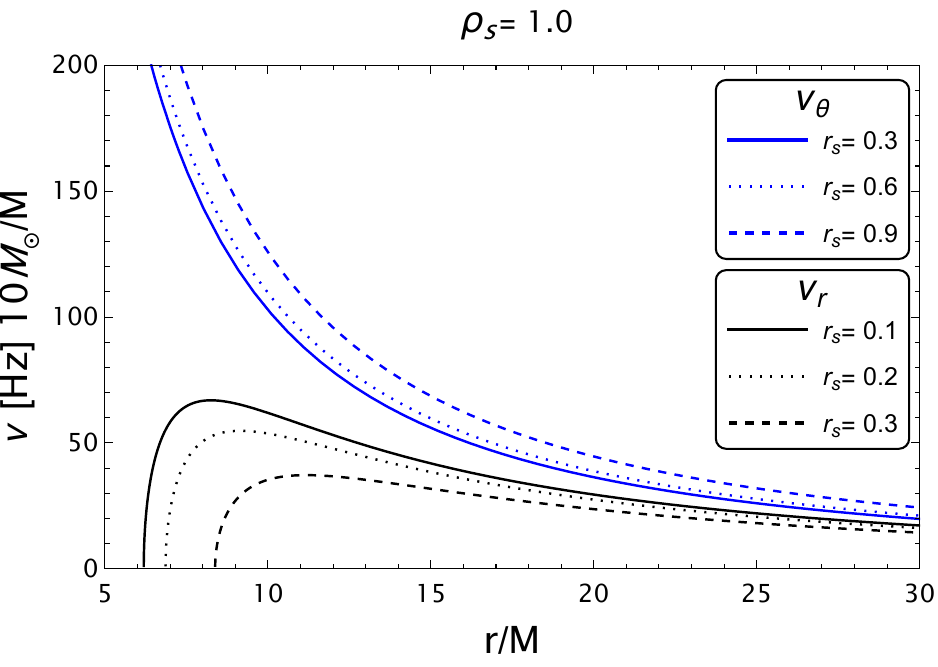}
    \includegraphics[scale=0.5]{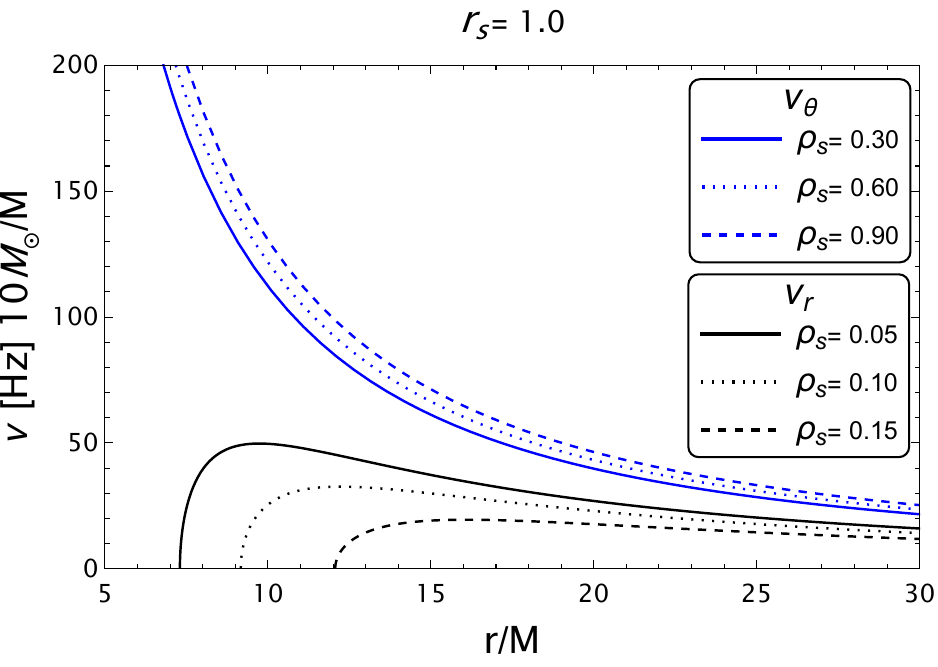}
    \caption{The radial profile of the frequencies $\nu_\theta$ and $\nu_r$ measured by a distant observer is shown as a function of $r/M$ for different values of $\rho_s$ and $r_s$. 
 }
    \label{fig:frequency}
\end{figure*}

Here, we consider a test particle motion moving in a stable circular orbit around a Schwarzschild-like black hole immersed in a Dehnen-type dark matter halo. We note that such a particle on a stable circular orbit begins to oscillate around it due to perturbations, and its motion is characterized by the epicyclic frequencies, which can be determined as follows \cite{Shaymatov20egb,Stuchlik21_qpo,Kolos15qpo,AngeliniApj,KluzniakAcPPB,StuchlikAA}

\begin{equation}\label{radial-freq}
    \Omega_r^2 = \frac{\partial_r^2 H_{pot}(r,\theta)|_{r_c,\frac{\pi}{2}}}{g_{rr}} \ ,
\end{equation}
\begin{equation} \label{latitudal-freq}
    \Omega_\theta^2 = \frac{\partial_\theta^2 H_{pot}(r,\theta)|_{r_c,\frac{\pi}{2}}}{g_{\theta\theta}} \ .
\end{equation}
Here, $\Omega_r$ and $\Omega_\theta$ denote the local epicyclic frequencies. However, since many astronomical sources are located at large distances, the epicyclic frequencies as measured by a distant observer are of greater physical relevance than the local ones. The relation between the local and distant epicyclic frequencies is expressed by the following equation:
\begin{equation}
    \omega_{\text{distant}} = \frac{\Omega_{\text{local}}}{u^t} \ ,
\end{equation}
Thus, the epicyclic frequencies as measured by a distant observer are given by the following expressions:
\begin{align} \label{radial-freq-exp-distant}
    &\omega_r^2 =\frac{M}{r^3}-\frac{6M^2}{r^4}-\frac{8 \pi  \rho_s r_s^2 \left(4 r^2+18 r r_s+15 r_s^2\right)M}{r^5 \left(\frac{r+r_s}{r}\right)^{3/2}} \nonumber \\ 
    &+\frac{4 \pi  \rho_s r_s^3 \left(-160 \pi  \rho_s r_s^3 -64 \pi  r \rho_s r_s^2 +\frac{2 r+3 r_s}{\sqrt{\frac{r+r_s}{r}}}\right)}{r^3 \left(r+r_s\right)}  \ ,
\end{align}
\begin{equation} \label{latitudal-freq-exp-distant}
    \omega_\theta^2 =\frac{M}{r^3}+\frac{8\pi\rho_s r^3_s}{r^3}\sqrt{\frac{r}{r+r_s}} \ .
\end{equation}
To convert the frequency to SI units (Hz), the following transformation is applied:
\begin{equation}
    \nu_i = \frac{\omega_i}{2\pi} \frac{c^3}{GM} \ .
\end{equation}
Fig.~\ref{fig:frequency} shows the epicyclic frequencies $\nu_r$ and $\nu_\theta$ as functions of $r/M$, as measured by a distant observer, for various values of the dark matter halo density $\rho_s$ and halo radius $r_s$. From the plots, it is evident that both $\rho_s$ and $r_s$ influence the epicyclic frequencies in a similar manner. Specifically, as $\rho_s$ and $r_s$ increase, the $\theta$-component of the epicyclic frequency, $\nu_\theta$, increases, while the radial component, $\nu_r$, decreases and shifts to larger radii.

This behavior can be interpreted as the result of an enhanced gravitational pull caused by the increase in $\rho_s$ and $r_s$, effectively mimicking the presence of additional mass at the center of the black hole. In other words, the gravitational field becomes stronger. Thus, the rise in $\nu_\theta$ results from the growth of the second term in Eq.~\eqref{latitudal-freq-exp-distant}. Additionally, the stronger gravitational field pushes the innermost stable circular orbit (ISCO) outward, causing radial oscillations to occur at larger radii. As a result, the radial epicyclic frequency $\nu_r$ shifts to the right and decreases in magnitude. It must be emphasized that the present analyses of time-like geodesics and the QPOs arising from the epicyclic motion of particles around black holes embedded in a Dehnen-type dark matter halo are of significant astrophysical importance. Similarly, previous studies within different dark matter models also showed that dark matter can influence epicyclic frequencies and even mimic the effects of black hole rotation (see, e.g., \cite{Stuchlik2022ApJ,Xamidov25PDU}).


\begin{figure*}
    \centering
\includegraphics[scale=0.35]{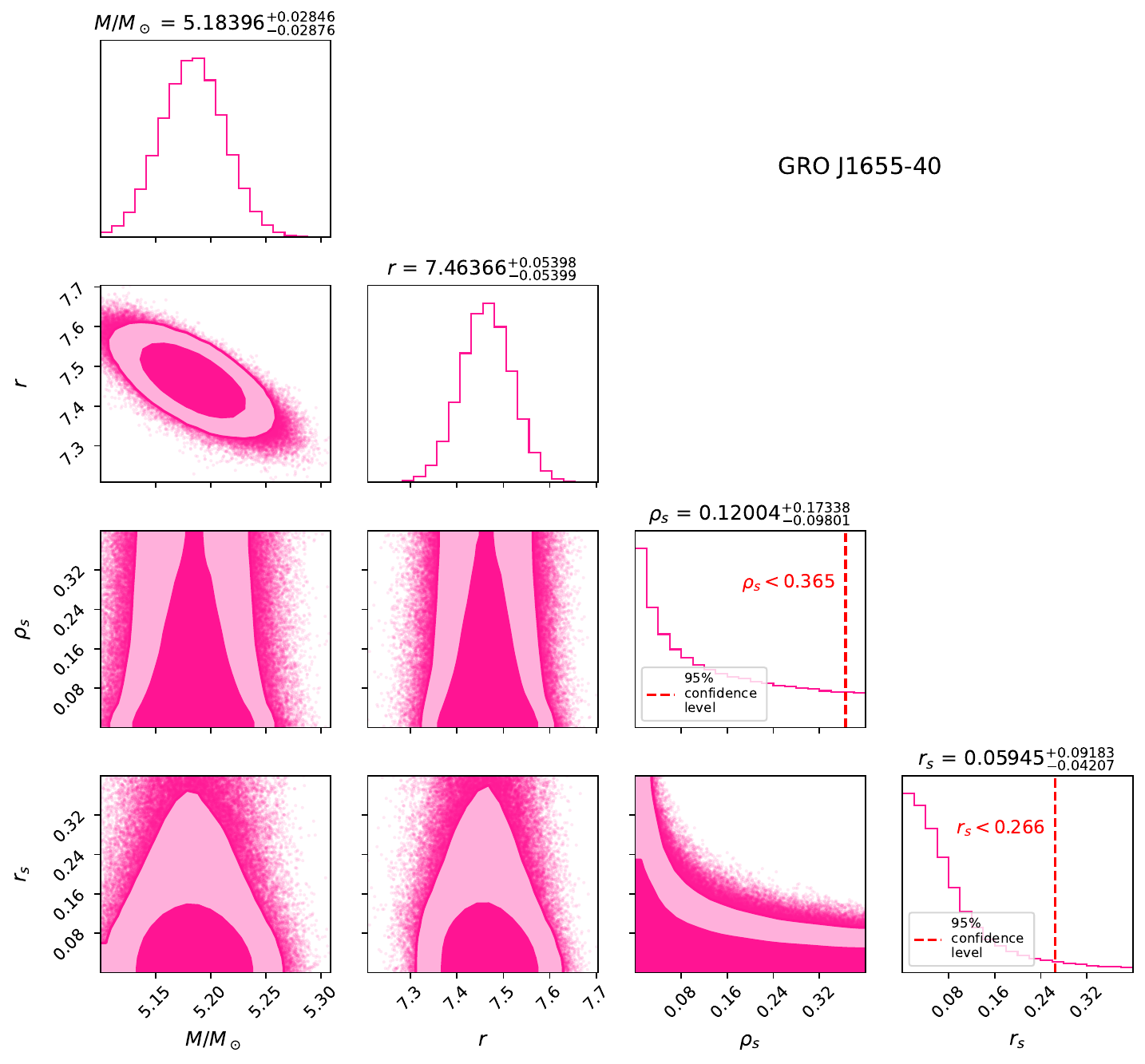}
\includegraphics[scale=0.35]{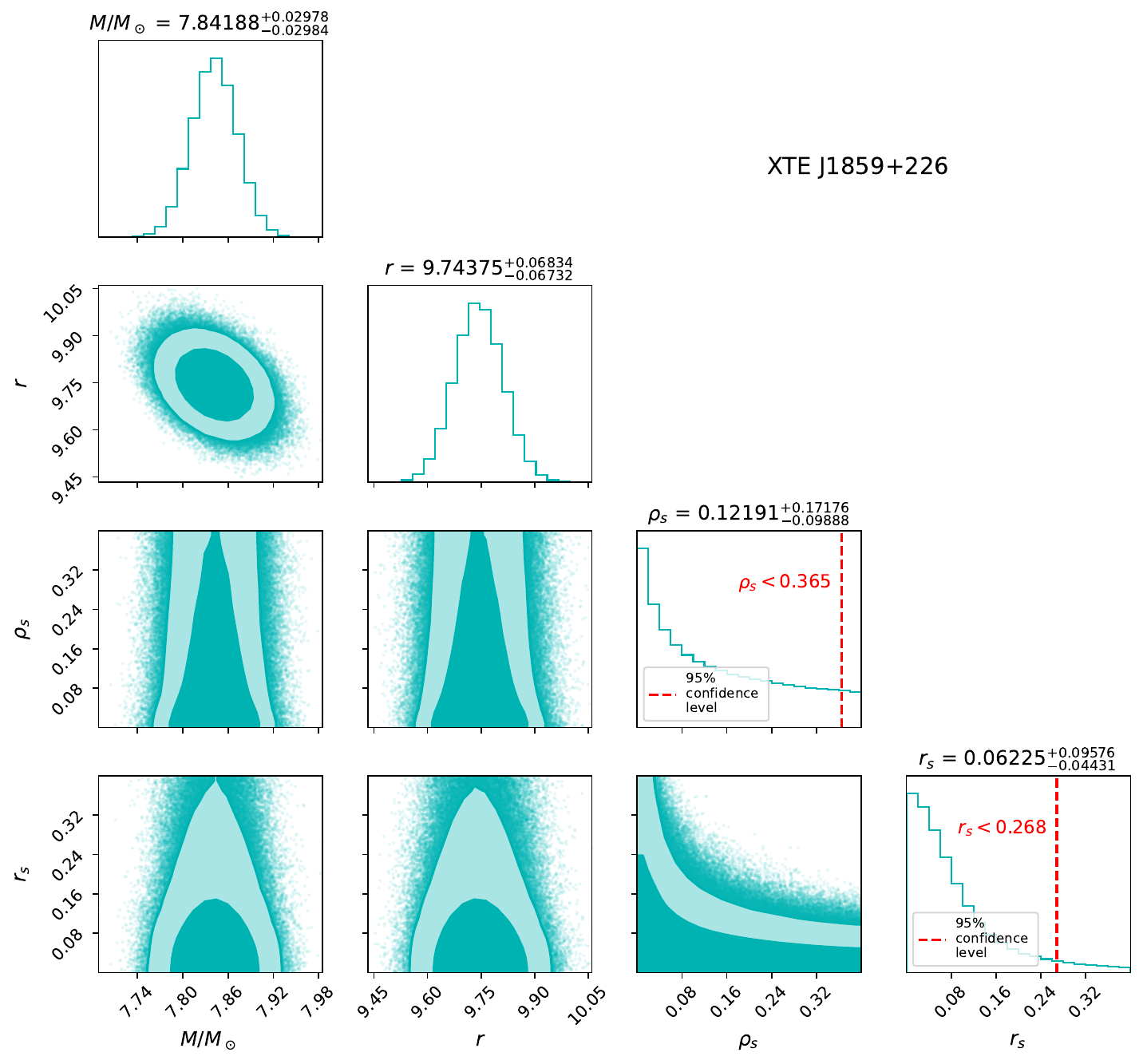}
\includegraphics[scale=0.35]{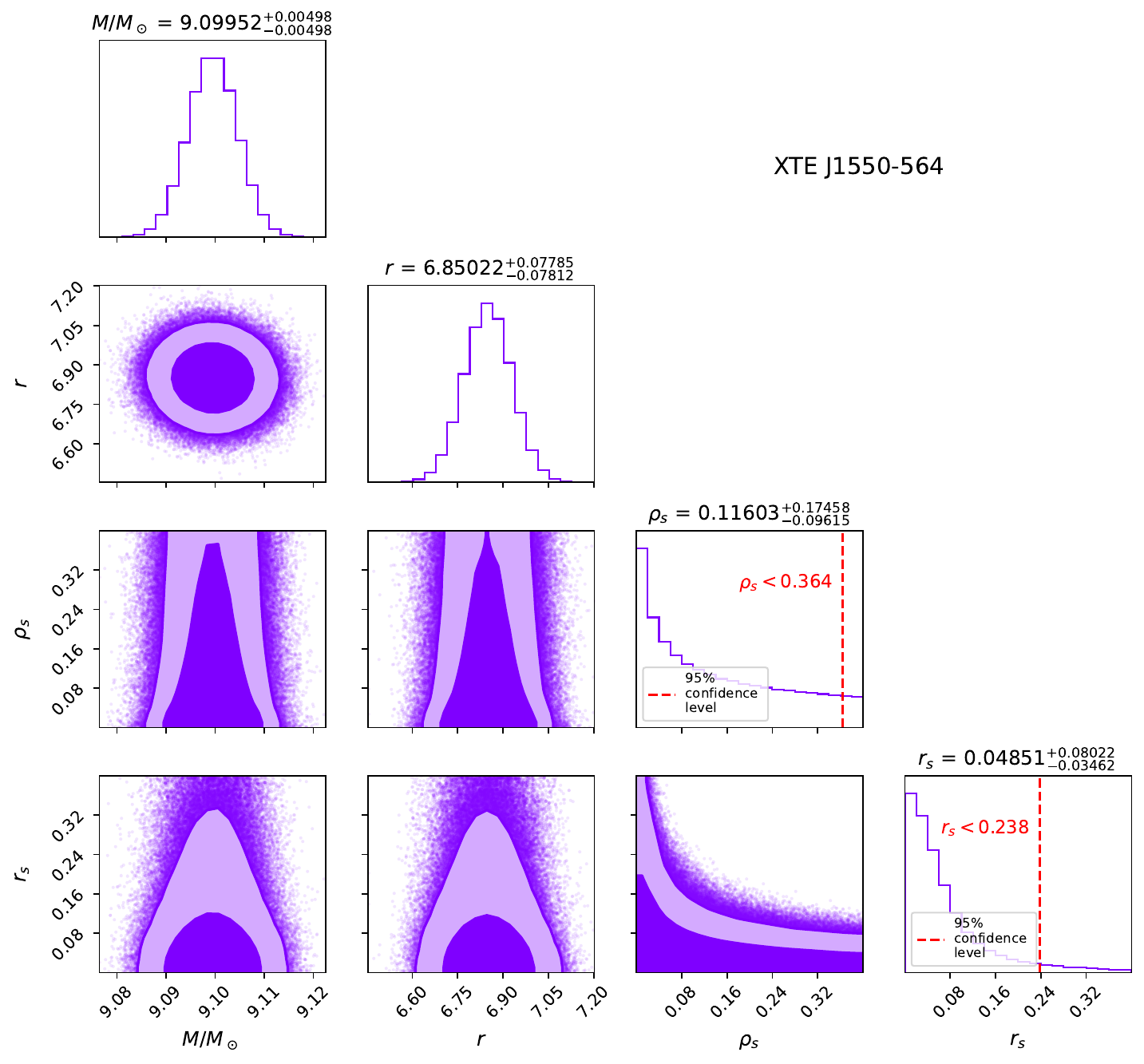}
\includegraphics[scale=0.35]{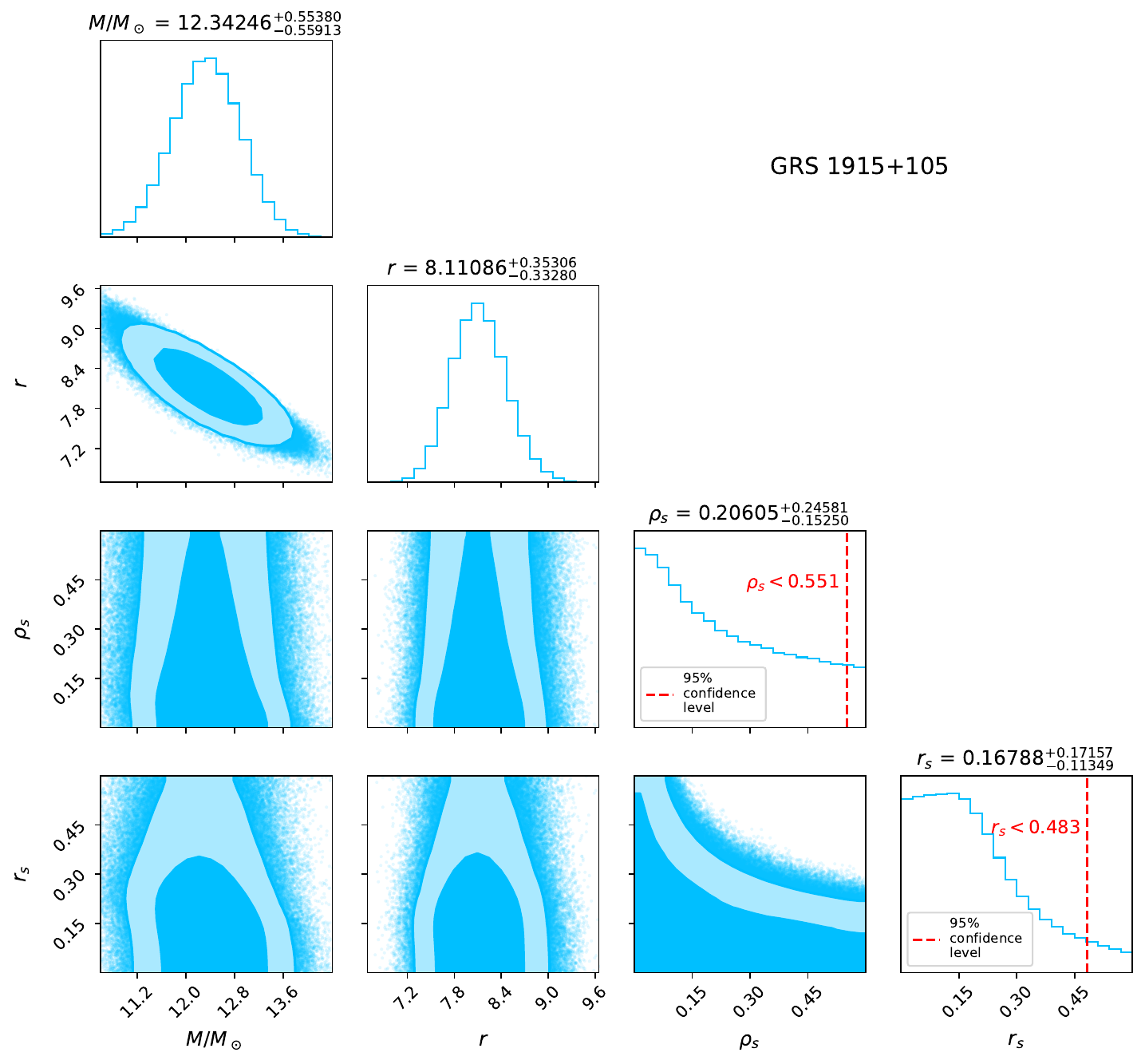}    
    \caption{ Posterior probability distributions for the black hole mass ($M$), the dimensionless QPO radius ($r/M$), and the dark matter halo density ($\rho_s$) and halo radius ($r_s$) - obtained with the FR model. Vertical red dashed lines mark the 95 \% credible upper limits on $\rho_s$ and $r_s$.}
    \label{Fig.mcmc}
\end{figure*}

\begin{figure*}
    \centering
\includegraphics[scale=0.5]{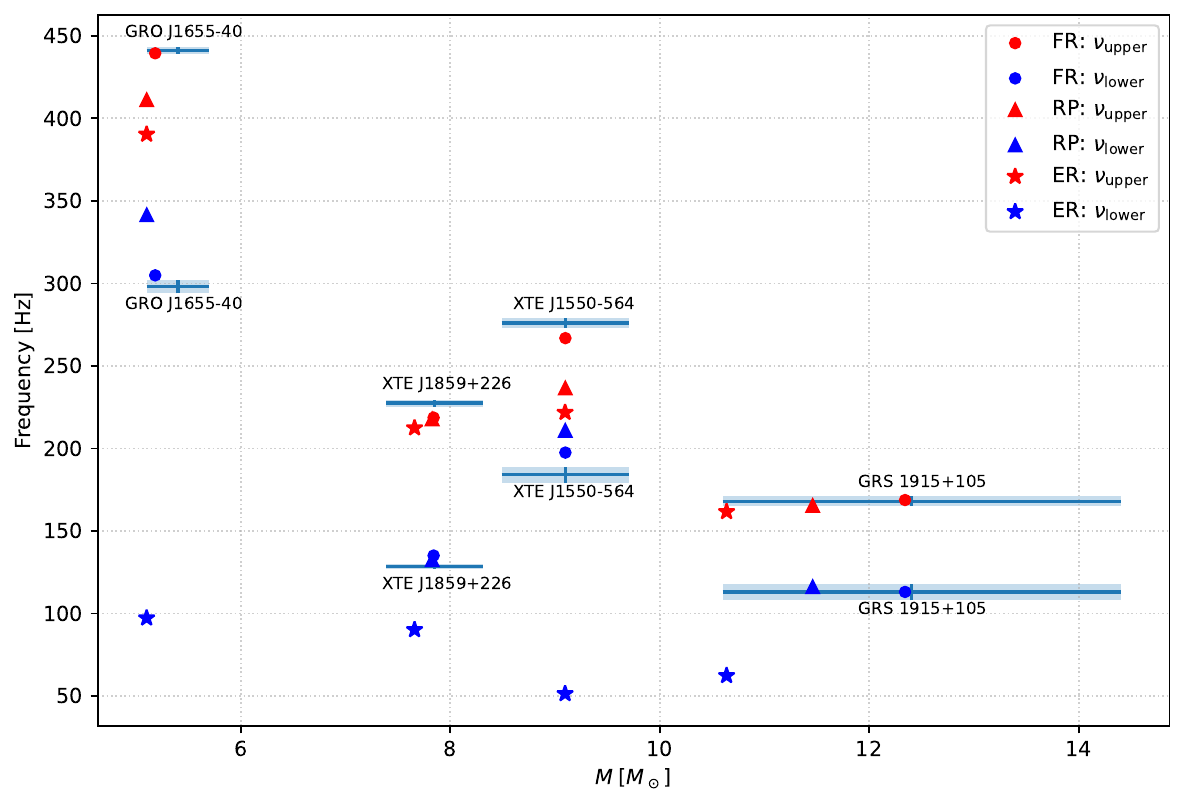}
    \caption{Comparison between model predictions and observational data for HF QPOs in X-ray binary systems. Dots, triangles and stars represent the model frequencies (upper and lower, respectively) calculated using best-fit parameters obtained from MCMC analysis (see Table~\ref{table:best-fit2}). The blue error boxes show the corresponding observational data, including uncertainties in both mass and QPO frequency (see Table~\ref{Table 1}). Each source is labeled accordingly.}
    \label{Fig.bestFit}
\end{figure*}

\section{Constrains on the parameters of the dark matter halo through the astrophysical quasiperiodic oscillations }\label{Sec:MCMC}

In this section, we constrain the dark matter halo density $\rho_s$ and halo radius $r_s$ using MCMC analysis based on observational data from four QPO sources. The validated  observational data for these sources are listed in Table~\ref{Table 1}. 
To describe the upper and lower QPO frequencies, we consider three geodesic QPO models --- forced resonance (FR), relativistic precession (RP), and epicyclic resonance (ER) --- with their respective frequency relations summarized in Table~\ref{tab:models}.

\begin{table}[]
\begin{tabular}{|l|c|c|}
\hline
\multicolumn{1}{|c|}{Model}  & \begin{tabular}[c]{@{}c@{}}Upper\\ frequency $\nu_{up}$\end{tabular} & \begin{tabular}[c]{@{}c@{}}Lower\\ frequency $\nu_{low}$\end{tabular} \\ \hline
Forced Resonance (FR)        & $\nu_{up} = \nu_\theta + \nu_r$                                      & $\nu_{low} = \nu_\theta$                                              \\ \hline
Relativistic Precession (RP) & $\nu_{up} = \nu_\theta$                                              & $\nu_{low} = \nu_\theta - \nu_r$                                      \\ \hline
Epicyclic Resonance (ER)     & $\nu_{up} = \nu_\theta$                                              & $\nu_{low} = \nu_r$                                                   \\ \hline
\end{tabular}
\caption{Geodesic QPO models applied to the observed twin HF QPOs~\cite{Deligianni2021,Banerjee_2022,Shaymatov23ApJ,Stuchlik2016AA}.}
\label{tab:models}
\end{table}

We use the Python package the \textit{emcee}, which implements Markov Chain Monte Carlo (MCMC) sampling, to find the best-fit values of the dark matter halo density $\rho_s$ and scale radius $r_s$~\cite{Foreman_Mackey_2013}.
Bayes' theorem yields the posterior probability distribution
\[
P(\Theta \mid D, \mathcal{M}) \;=\; 
\frac{\mathcal{L}(D \mid \Theta, \mathcal{M}) \,
      \pi(\Theta \mid \mathcal{M})}
     {P(D \mid \mathcal{M})}\,,
\]
where
$P(\Theta \mid D, \mathcal{M})$ is the \textit{posterior},
$\mathcal{L}(D \mid \Theta, \mathcal{M})$ is the \textit{likelihood},
$\pi(\Theta \mid \mathcal{M})$ is the \textit{prior},
and $P(D \mid \mathcal{M})$ is a \textit{normalization constant}.

\renewcommand{\arraystretch}{1.5}
\begin{table}[]
\resizebox{.5\textwidth}{!}{
\begin{tabular}{|c|c|c|c|c|}
\hline
        & GRS 1915 + 105 \cite{Remillard_2006} & XTE J1859+226  \cite{Molla_2017,Ingram_2014}      & XTE J1550 - 564 \cite{Remillard_2002,Orosz_2011} & GRO J1655 - 40 \cite{mota2013} \\ \hline
$M [M_{\odot}]$    & $12.4^{+2.0}_{-1.8} $   & $7.85^{+0.46}_{-0.46} $  & $9.1^{+0.61}_{-0.61}$       & $5.4^{+0.3}_{-0.3}$        \\ \hline
$\nu_U [Hz]$  & $168^{+3.0}_{-3.0} $      & $227.5^{+2.1}_{-2.4} $         & $276^{+3.0}_{-3.0}$          & $441^{+2.0}_{-2.0}$         \\ \hline
$\nu_L [Hz]$ & $113^{+5.0}_{-5.0} $        & $128.6^{+1.6}_{-1.8}$         & $184^{+5.0}_{-5.0}$          & $298^{+4.0}_{-4.0}$         \\ \hline
\end{tabular}
}
 \caption{Masses and corresponding the upper and the lower QPO frequencies for the X-ray binaries analyzed in this work.}
    \label{Table 1}
\end{table}

We assign a Gaussian prior to the mass of the black hole $M$,
\begin{equation}
    \pi(M)\;=\;
\frac{1}{\sqrt{2\pi}\,\sigma}\,
\exp\!\left[-\frac{(M-M_0)^2}{2\sigma^2}\right]\,,
\end{equation}
$M_{\min} < M < M_{\max},$ and \(\pi(M)=0\) outside that interval. Here $M_{\min}$ and $M_{\max}$ denote the lower and upper bounds of the black hole mass, taken from the observational uncertainties reported in Table~\ref{Table 1} for each X-ray binary.

For the remaining parameters \(\theta_i = [r,\;\rho_s,\;r_s]\) we use
independent uniform (“top–hat”) priors,
\[
\pi(\theta_i)=
\begin{cases}
(\theta_{i,\max}-\theta_{i,\min})^{-1}, & \theta_{i,\min}<\theta_i<\theta_{i,\max},\\[4pt]
0, & \text{otherwise}.
\end{cases}
\]
The numerical limits \(\theta_{i,\min}\) and \(\theta_{i,\max}\)
are given in Table~\ref{table:priors}.

Assuming independence, the full prior factorises as
\[
\pi(\Theta)\;=\;
\pi(M)\,\pi(r)\,\pi(\rho_s)\,\pi(r_s),
\]
where \(\Theta=(M,r,\rho_s,r_s)\).

\renewcommand{\arraystretch}{1.5}
\begin{table}[]
\resizebox{.5\textwidth}{!}{
\begin{tabular}{|c|cc|cc|cc|cc|}
\hline
\multirow{2}{*}{Parameters} & \multicolumn{2}{c|}{GRS 1915 + 105}   & \multicolumn{2}{c|}{XTE J1859+226}        & \multicolumn{2}{c|}{XTE J1550 - 564}  & \multicolumn{2}{c|}{GRO J1655 - 40}   \\ \cline{2-9} 
                            & \multicolumn{1}{c|}{$\ \quad\mu\quad\ $} & $\sigma$ & \multicolumn{1}{c|}{$\ \quad\mu\quad\ $} & $\sigma$ & \multicolumn{1}{c|}{$\ \quad\mu\quad\ $} & $\sigma$ & \multicolumn{1}{c|}{$\ \quad\mu\quad\ $} & $\sigma$ \\ \hline
$M/M_{\odot}$               & \multicolumn{1}{c|}{12.4}  & 0.6      & \multicolumn{1}{c|}{7.85} & 0.03      & \multicolumn{1}{c|}{9.1}   & 0.005      & \multicolumn{1}{c|}{5.2}   & 0.03      \\ \hline
$r/M$                     & \multicolumn{8}{c|}{Uniform (6, 15)} 
\\ \hline
$\rho_s/M^2$                     & \multicolumn{2}{c|}{Uniform [0, 0.6)} & \multicolumn{2}{c|}{Uniform [0, 0.4)} & \multicolumn{2}{c|}{Uniform [0, 0.4)} & \multicolumn{2}{c|}{Uniform [0, 0.4)} \\ \hline
$r_s/M$                     & \multicolumn{2}{c|}{Uniform [0, 0.6)} & \multicolumn{2}{c|}{Uniform [0, 0.4)} & \multicolumn{2}{c|}{Uniform [0, 0.4)} & \multicolumn{2}{c|}{Uniform [0, 0.4)} \\ \hline
\end{tabular}
}
\caption{Prior distributions used in the MCMC analysis: Gaussian priors for the masses ($\mu$ – mean, $\sigma$ – standard deviation) and uniform priors for the orbital radius $r/M$ as well as for the dark matter halo density $\rho_s/M^{2}$ and halo radius $r_s/M$ - for each X-ray binary.}
\label{table:priors}
\end{table}

Based on the upper and lower frequency relations given in Table~\ref{tab:models}, we construct a joint Gaussian likelihood that simultaneously incorporates the observational $\nu_{\mathrm{up}}$ and $\nu_{\mathrm{low}}$ data listed in Table~\ref{Table 1}.  The total log-likelihood is
\begin{equation}
\ln\mathcal{L} \;=\; \ln\mathcal{L}_{U} \;+\; \ln\mathcal{L}_{L}\,,
\end{equation}
where the contribution from the upper (\(U\)) frequencies is
\begin{equation}
\ln\mathcal{L}_{U} \;=\;
-\frac12
\sum_{i}
\frac{\bigl(\nu_{U,\mathrm{obs}}^{\,i}-\nu_{U,\mathrm{mod}}^{\,i}\bigr)^{2}}
     {(\sigma_{U,\mathrm{obs}}^{\,i})^{2}}
\,,
\end{equation}

and the term for the lower (\(L\)) frequencies is

\begin{equation}
\ln\mathcal{L}_{L} \;=\;
-\frac12
\sum_{i}
\frac{\bigl(\nu_{L,\mathrm{obs}}^{\,i}-\nu_{L,\mathrm{mod}}^{\,i}\bigr)^{2}}
     {(\sigma_{L,\mathrm{obs}}^{\,i})^{2}}
\,.
\end{equation}

Here,
\(\nu_{U,\mathrm{obs}}^{\,i}\) and \(\nu_{L,\mathrm{obs}}^{\,i}\) are the measured upper and lower QPO frequencies,
\(\nu_{U,\mathrm{mod}}^{\,i}\) and \(\nu_{L,\mathrm{mod}}^{\,i}\) are the corresponding model predictions, and
\(\sigma_{U,\mathrm{obs}}^{\,i}\) and \(\sigma_{L,\mathrm{obs}}^{\,i}\) denote
the corresponding statistical uncertainties for the associated quantities.

With this likelihood we explore the four-parameter space
$[r, M, \rho_s, r_s]$ via MCMC.
The best-fit values and upper limits of the black hole parameters obtained with different QPO models are reported in Table~\ref{table:best-fit2}, while Fig.~\ref{Fig.mcmc} represents the posterior probability density distributions for the four X-ray binaries in the case of the FR model.
The shaded regions in Fig.~\ref{Fig.mcmc} mark the 68\,\%, and 95\,\% confidence level, illustrating the statistical uncertainties of the inferred dark matter halo parameters $\rho_s$ and $r_s$.

The agreement between model predictions and observations can be assessed using the following expression
\begin{eqnarray}
    \chi^{2}
&=& \frac{\big(f_{\mathrm{up}}^{\mathrm{obs}} - f_{\mathrm{up}}^{\mathrm{model}}\big)^{2}}{\sigma_{\mathrm{up}}^{2}}
+ \frac{\big(f_{\mathrm{low}}^{\mathrm{obs}} - f_{\mathrm{low}}^{\mathrm{model}}\big)^{2}}{\sigma_{\mathrm{low}}^{2}}\, \\ \nonumber
&+& \frac{\big(M^{\mathrm{obs}} - M^{\mathrm{model}}\big)^{2}}{\sigma_{M}^{2}} \, .
\end{eqnarray}

This model considered here provides a good fit when the corresponding reduced $\chi^2$ approaches unity, indicating that deviations between model predictions and observations are consistent with the observational uncertainties.

The calculated $\chi^2$ values for the four X-ray binaries under each model are summarized in Table~\ref{tab:chi2_models}, and Fig.~\ref{Fig.bestFit} presents a comparison between the model predictions and the observational data.
From the $\chi^{2}$ analysis in Table~\ref{tab:chi2_models} and Fig.~\ref{Fig.bestFit}, the FR model provides the best fit for GRS~1915+105, whereas the RP model exhibits noticeable deviations from the observational data. For GRO~J1655-40, the FR model yields the lowest $\chi^{2}$ value among the considered models, although some discrepancies with the data remain.

\begin{table}[]
\resizebox{.5\textwidth}{!}{
\begin{tabular}{|c|c|cccc|cc|l}
\cline{1-8}
\multirow{2}{*}{Model} & \multirow{2}{*}{Object} & \multicolumn{4}{c|}{Best-fit}                                                                               & \multicolumn{2}{c|}{Upper limit}                         &  \\ \cline{3-8}
                       &                         & \multicolumn{1}{c|}{$M/M_\odot$} & \multicolumn{1}{c|}{$r/M$} & \multicolumn{1}{c|}{$\rho_s/M^2$} & $r_s/M$ & \multicolumn{1}{c|}{$\rho_s/M^2$}     & $r_s/M$          &  \\ \cline{1-8}
\multirow{4}{*}{FR}    & GRO J1655-40            & \multicolumn{1}{c|}{5.184}       & \multicolumn{1}{c|}{7.464} & \multicolumn{1}{c|}{0.120}        & 0.059   & \multicolumn{1}{c|}{\textless{}0.365} & \textless{}0.266 &  \\ \cline{2-8}
                       & XTE J1859+226           & \multicolumn{1}{c|}{7.842}       & \multicolumn{1}{c|}{9.744} & \multicolumn{1}{c|}{0.122}        & 0.062   & \multicolumn{1}{c|}{\textless{}0.365} & \textless{}0.268 &  \\ \cline{2-8}
                       & XTE J1550-564           & \multicolumn{1}{c|}{9.099}       & \multicolumn{1}{c|}{6.850} & \multicolumn{1}{c|}{0.116}        & 0.049   & \multicolumn{1}{c|}{\textless{}0.364} & \textless{}0.238 &  \\ \cline{2-8}
                       & GRS 1915+105            & \multicolumn{1}{c|}{12.342}      & \multicolumn{1}{c|}{8.111} & \multicolumn{1}{c|}{0.206}        & 0.168   & \multicolumn{1}{c|}{\textless{}0.551} & \textless{}0.483 &  \\ \cline{1-8}
\multirow{4}{*}{RP}    & GRO J1655-40            & \multicolumn{1}{c|}{5.102}       & \multicolumn{1}{c|}{6.178} & \multicolumn{1}{c|}{0.106}        & 0.016   & \multicolumn{1}{c|}{\textless{}0.362} & \textless{}0.116 &  \\ \cline{2-8}
                       & XTE J1859+226           & \multicolumn{1}{c|}{7.827}       & \multicolumn{1}{c|}{7.094} & \multicolumn{1}{c|}{0.117}        & 0.045   & \multicolumn{1}{c|}{\textless{}0.364} & \textless{}0.231 &  \\ \cline{2-8}
                       & XTE J1550-564           & \multicolumn{1}{c|}{9.098}       & \multicolumn{1}{c|}{6.073} & \multicolumn{1}{c|}{0.109}        & 0.025   & \multicolumn{1}{c|}{\textless{}0.363} & \textless{}0.169 &  \\ \cline{2-8}
                       & GRS 1915+105            & \multicolumn{1}{c|}{11.460}      & \multicolumn{1}{c|}{6.613} & \multicolumn{1}{c|}{0.181}        & 0.090   & \multicolumn{1}{c|}{\textless{}0.547} & \textless{}0.394 &  \\ \cline{1-8}
\multirow{4}{*}{ER}    & GRO J1655-40            & \multicolumn{1}{c|}{5.101}       & \multicolumn{1}{c|}{6.397} & \multicolumn{1}{c|}{0.102}        & 0.009   & \multicolumn{1}{c|}{\textless{}0.361} & \textless{}0.082 &  \\ \cline{2-8}
                       & XTE J1859+226           & \multicolumn{1}{c|}{7.659}       & \multicolumn{1}{c|}{7.319} & \multicolumn{1}{c|}{0.103}        & 0.017   & \multicolumn{1}{c|}{\textless{}0.362} & \textless{}0.131 &  \\ \cline{2-8}
                       & XTE J1550-564           & \multicolumn{1}{c|}{9.096}       & \multicolumn{1}{c|}{6.341} & \multicolumn{1}{c|}{0.105}        & 0.018   & \multicolumn{1}{c|}{\textless{}0.362} & \textless{}0.139 &  \\ \cline{2-8}
                       & GRS 1915+105            & \multicolumn{1}{c|}{10.638}      & \multicolumn{1}{c|}{7.053} & \multicolumn{1}{c|}{0.161}        & 0.035   & \multicolumn{1}{c|}{\textless{}0.544} & \textless{}0.240 &  \\ \cline{1-8}
\end{tabular}
}
\caption{Best-fit values and upper limits for the parameters of the Schwarzschild black hole immersed in a dark matter halo for various QPO models.}
\label{table:best-fit2}
\end{table}

\begin{table}[h!]
\centering
\begin{tabular}{|c|c|c|c|c|}
\hline
Object        & $\chi^2_{\mathrm{RP}}$ & $\chi^2_{\mathrm{ER}}$ & $\chi^2_{\mathrm{FR}}$ & Best Model \\
\hline
GRO J1655-40  & 341.638 & 3163.086 & 4.107  & FR \\
\hline
XTE J1550-564 & 201.530 & 1030.014 & 16.655 & FR \\
\hline
XTE J1859+226 & 27.134  & 629.288  & 34.149 & RP \\
\hline
GRS 1915+105  & 1.443   & 107.897  & 0.057  & FR \\
\hline
\end{tabular}
\caption{Comparison of $\chi^2$ values for different models across the considered objects, with the best-fitting model reported in the last column.}

\label{tab:chi2_models}
\end{table}

\section{Conclusions}
\label{Sec:conclusion}

In this work, we examined a Schwarzschild BH embedded in a Dehnen-type $(1,4,5/2)$ DM halo. By combining  theory with observational constraints, we showed that the DM halo can alter the orbits of objects measurably. A perihelion shift expression, applied to Mercury and the orbit of the S2 star, reveals that the admissible $(\rho_s,r_s)$ parameter space broadens with increasing mass of the BH, implying that DM halo effects are more evident around more massive orbits. Our findings show that the effects of the DM halo become observable only around supermassive BHs.

Our analysis of QPOs shows that epicyclic frequencies can carry a clear halo signature and indicates that increasing either the halo radius $r_s$ or density $\rho_s$ raises the vertical (latitudinal) epicyclic frequency $\nu_\theta$, while it lowers the radial frequency $\nu_r$. This behavior explained by the extra gravitational pull of the DM halo, which shifts the innermost stable circular orbit ($r_{\text{ISCO}}$) to larger radii.

Twin-QPO measurements place empirical limits on the halo. To constrain the halo parameters empirically, we perform MCMC analysis with the \textit{emcee} sampler \cite{Foreman_Mackey_2013}. Figure \ref{Fig.mcmc} presents the posterior distributions for four Galactic microquasars, including GRO J1655-40, GRS 1915+105, XTE J1859+226, and XTE J1550-564. The corresponding best-fit and upper limit values are listed in Table \ref{table:best-fit2}. Using these data, we computed model QPO frequencies and compared them with the observed values, as shown in Fig.~\ref{Fig.bestFit}. 
We showed that the FR model provides the best fit for GRS~1915+105 and yields reasonably good agreement for GRO~J1655-40, while noticeable discrepancies remain for XTE~J1550-564 and XTE~J1859+226.
Our combined geodesic analysis and constraints using the data of QPOs demonstrate that timelike orbits and epicyclic oscillations around a BH can serve as sensitive probes of the surrounding DM halo. Overall, our results enhance our understanding of dark matter and enable us to discriminate the Dehnen-type DM halo from alternative halo profiles, which could be crucial for identifying DM signatures in future analysis and astronomical observations.

\acknowledgments

We are grateful to the anonymous referee for useful comments and constructive suggestions that definitely helped us improve the clarity and quality of the manuscript. S.S. is supported by the National Natural Science Foundation of China under Grant No. W2433018. T. Zhu is also supported by the National Natural Science Foundation of China under Grants No. 12275238, the National Key Research and Development Program of China under Grant No. 2020YFC2201503, and the Zhejiang Provincial Natural Science Foundation of China under Grants No. LR21A050001 and No. LY20A050002.

%
\bibliographystyle{apsrev4-1}  
\bibliography{Ref1,Ref2}

\end{document}